# The Origin of Ti *1s* XANES Main Edge Shifts and EXAFS Oscillations in the Energy Storage Materials Ti₂CTₓ and Ti₃C₂Tₓ MXenes

## Lars-Åke Näslund[1] and Martin Magnuson[1]

[1] Thin Film Physics Division, Department of Physics, Chemistry, and Biology (IFM), Linköping University, SE-581 83 Linköping, Sweden.

E-mail: lars-ake.naslund@liu.se

## Abstract

A potential application of two-dimensional (2D) MXenes, such as $Ti_2CT_x$ and $Ti_3C_2T_x$, is energy storage devices, such as supercapacitors, batteries, and hydride electrochemical cells, where intercalation of ions between the 2D layers is considered as a charge carrier. Electrochemical cycling investigations in combination with Ti *1s* X-ray absorption spectroscopy (XAS) have therefore been performed with the objective to study oxidation state changes during potential variations. In some of these studies $Ti_3C_2T_x$ has shown main edge shifts in the Ti *1s* X-ray absorption near-edge structure (XANES). Here we show that these main edge shifts originate from the Ti *4p* orbital involvement in the bonding between the surface Ti and the termination species at the fcc-sites. The study further shows that the $t_{2g}$-$e_g$ crystal field splitting (10Dq) observed in the pre-edge absorption region indicate weaker Ti-C bonds in $Ti_2CT_x$ and $Ti_3C_2T_x$ compared to TiC and the corresponding MAX phases. The results from this study provide information necessary for improved electronic modeling and subsequently a better description of the materials properties of the MXenes. In general, potential applications, where surface interactions with intercalation elements are important processes, will benefit from the new knowledge presented.

Keywords: X-ray absorption spectroscopy, near edge structures, $t_{2g}$-$e_g$ crystal field splitting, Ti-C bond, bond strength, bond length, electrochemical cycling.

## 1. Introduction

Transition metal carbides or nitrides known as MXenes ($M_{n+1}X_nT_x$, n = 1, 2, 3) have shown promising results in many different applications [1-6]. Most attention is dedicated to MXenes as energy storage materials [1,2]. For example, $Ti_3C_2T_x$ in acidic aqueous electrolytes has provided a high volumetric capacitance and decent rate capability [7]. Hence, $Ti_3C_2T_x$ shows promising results as a charge storage material in rechargeable lithium (Li) or sodium (Na) ion batteries and supercapacitors [8-11]. The promising charge storage capability of $Ti_3C_2T_x$ has been attributed to the fast pseudocapacitive ion intercalation/deintercalation process with a surface redox reaction where the charge carrier adsorbs on the $Ti_3C_2T_x$ surface [7,12]. It is therefore highly motivated to study this process in detail, as electrochemical energy storage is a key feature in future high energy density devices.

A powerful tool employed to facilitate the development of electrochemical energy storage materials is *K*-edge X-ray absorption near-edge structure (XANES), because the technique is sensitive to oxidation state changes caused by modifications in the chemical surroundings and the local bonding structure around the probed element. In addition, $K$-edge XANES of transition metals require photons in the hard X-ray regime, which makes the technique suitable for in situ studies of the Ti oxidation state changes during ion intercalation of, e.g., $Ti_3C_2T_x$ [12-17].

However, a recent XANES-study of $Ti_3C_2T_x$ showed that the main-edge absorption features of the Ti $1s$ XANES, which mainly originate from the Ti $1s \rightarrow 4p$ excitations, are sensitive towards the fcc-site occupation [18]. The finding raises the question whether it is an oxidation of the Ti atoms that shifts the main edge in the intercalation process, as implied in previous electrochemical studies [12-17], or an alteration of the species in the fcc-site that causes the main-edge shift. A deeper study is therefore motivated where the origin of the Ti $1s$ XANES main-edge shift is clarified. Therefore, in this work, we elucidate the origin of the main edge shift and how it is controlled by the local structural properties and interactions around the Ti atoms in TiC, $Ti_2CT_x$ and $Ti_3C_2T_x$ as well as the parent compounds $Ti_2AlC$ and $Ti_3AlC_2$ [19], where the selected components are assumed to have different Ti oxidation states [20] but otherwise many similarities. We will also scrutinize the effect of alternating the species on the fcc-site.

While $K$-edge XANES focuses on the fine structure near the edge of the X-ray absorption spectrum of the probed element, which explores the unoccupied density of states available for the $1s$ electron excitation, we have also expanded the energy range to include the extended X-ray absorption fine structure (EXAFS), which provides information about the chemical coordination and bond lengths around the probed element [21]. The surfaces of $Ti_2CT_x$ and $Ti_3C_2T_x$ are assumed to be nearly identical, although there is limited experimental support so far. Through EXAFS we explore similarities and dissimilarities between, e.g., the bond lengths of the termination species at the $Ti_2CT_x$ surface versus the $Ti_3C_2T_x$ surface.

MXenes are obtained from exfoliation of MAX phases, where the monolayers of A-atoms (A-layers) that separate the $M_{n+1}X_n$-layers are removed through selective etching [19]. Recently, a critical review of MAX phases from an electronic-structure and chemical bonding perspective was presented [22]. The review was discussing the special crystal structure and bonding characteristics in MAX phases and how they are related to their unique combination of metallic and ceramic features. The elements in the crystal structure of the MAX phases can be altered with the aim to induce variations in the conductivity, elasticity, magnetism and other material properties, with the purpose to tailor the material characteristics by controlling the strengths of their chemical bonds. This characteristic attribute is preserved in MXenes.

High photon energy $K$-edge X-ray absorption spectroscopy (XAS) is an excellent tool in revealing specific material characteristics. In the present work Ti $1s$ XAS has been employed to show differences and similarities regarding Ti-C distances in $Ti_2CT_x$ and $Ti_3C_2T_x$ in comparison with $Ti_2AlC$ and $Ti_3AlC_2$. A closer investigation of the crystal field splitting of the $t_{2g}$ and $e_g$ orbitals is also included, which reveals information about the Ti-C nucleus-nucleus repulsion, the strengthening and weakening of the covalent Ti-C bonds, and how the termination species affects the electron occupation in molecular orbitals that have both p- and d-character. An in situ heating treatment of the $Ti_3AlC_2$ sample will show if the Ti $1s$ XANES main edge is sensitive toward species coordinated to the fcc-sites on the MXenes, which then would require a Ti $4p$ involvement in the bonding between the Ti and the termination species. Hence, the study aims to provide information needed for improved electronic modeling and subsequently a better description of the materials properties of the MXenes.

## 2. Method

### 2.1 Sample preparation

Powders of TiC were obtained from a commercial supplier (Alfa Aesar, 98+%). Powders of $Ti_2AlC$ were obtained from a mixture of Ti (Alfa Aesar, 98+%), and Al (Alfa Aesar, 98+%) and graphite (Alfa Aesar, 99+%) of 2:1.1:1 molar ratios at 1400 °C and 240 min. Powders of $Ti_3AlC_2$ were obtained from a mixture of TiC (Alfa Aesar, 98+%), Ti (Alfa Aesar, 98+%), and Al (Alfa Aesar, 98+%) of 1:1:2 molar ratios at 1450 °C and 280 min. Formation of high-quality MAX phases were confirmed using Ni-filtered Cu Kα radiation in a normal Bragg−Brentano geometry of an X'Pert Panalytical X-ray diffraction (XRD) system [23].

A few mg of the fine TiC, $Ti_2AlC$ and $Ti_3AlC_2$ powders were mixed with polyethylene powder (Aldrich, 40-48 μm particle size) and thereafter cold pressed into ~500 μm thick pellets for the X-ray absorption spectroscopy.

Half a gram of each $Ti_2AlC$ and $Ti_3AlC_2$ powders were converted to $Ti_2CT_x$ and $Ti_3C_2T_x$, respectively, through selective etching of the Al-layers in 12 M HCl (Fisher, technical grade) and 2.3 M LiF (Alfa Aesar, 98+%). Details about the conversion process are presented in a previous publication [18]. The formations of MXenes were confirmed using XRD [23] on small pieces that were cut out from the obtained $Ti_2CT_x$ and $Ti_3C_2T_x$ freestanding films.

The obtained freestanding films of $Ti_2CT_x$ and $Ti_3C_2T_x$ were immediately stored in argon (Ar) atmosphere, except the small pieces for the XRD measurements. Mounting the $Ti_2CT_x$ and $Ti_3C_2T_x$ films on the sample holder were performed in a glove-bag filled with nitrogen gas ($N_2$).

### 2.2 XANES and EXAFS measurements

The MXenes samples were placed in a water-cooled gas cell (Linkam Scientific Instruments). A low $N_2$ flow protected the samples from oxidation during XAS. A heating element in the Linkam gas cell enabled XANES acquisition at different temperatures.

The Ti *1s* XAS spectra were recorded at BALDER, a wiggler beamline on the 3 GeV electron storage ring at MAX IV in Lund, Sweden. The Si(111) crystal monochromator provided an energy resolution of 1.0 eV and the photon energy scale was calibrated using the first derivative of the absorption spectrum from a Ti metal foil. The first inflection point was set to 4.9660 keV.

The XANES and EXAFS spectra were obtained in transmission mode. The photon beam intensity was monitored using an ionization chamber with 200 mbar $N_2$ and He gas mixture before the sample and an ionization chamber with 2 bar $N_2$ after the sample. Normalization of the XAS spectra was performed below the absorption edge. Thereafter the background intensity was subtracted before a second normalization above the absorption edge, i.e., in the photon energy region of 5.045-5.145 keV. Self-absorption effects were negligible in the Ti *1s* XANES spectra of the $Ti_2CT_x$ and $Ti_3C_2T_x$ freestanding films. The Ti *1s* XANES spectra of the $Ti_2AlC$, $Ti_3AlC_2$, and TiC pellets showed, on the other hand, self-absorption effects and have therefore been corrected using the simple function:

$$I_{corr}(\varepsilon) = AI(\varepsilon) \, [1\text{-}BI(\varepsilon)]^{-1} \qquad (1)$$

where A is a scaling factor, B is the self-absorption compensation factor, and $I(\varepsilon)$ is the Ti *1s* XANES spectrum. The coefficients A and B were adjusted until the absorption spectrum of the $Ti_2AlC$, $Ti_3AlC_2$, and TiC samples had similar appearance in the photon energy region of 5.045-5.145 keV as the corresponding spectra of $Ti_2CT_x$ and $Ti_3C_2T_x$.

*2.3 EXAFS analysis*

The fitting of the EXAFS was performed using the Visual Processing in EXAFS Researches (VIPER) software package [24], where the Ti-Ti, Ti-C, Ti-O and Ti-F scattering paths were obtained from the effective scattering amplitudes using the FEFF code [25,26]. The raw absorption data was accumulated from 10 absorption spectra and corrected from known monochromator-induced glitches. The background was subtracted before the normalization. The $k^2$-weighted $\chi$ EXAFS oscillations were thereafter derived.

A Hanning window function [25,26] with a many body factor of $S_0^2$=0.8 was employed to fit the back-Fourier-transform signal between k=0–15.5 Å$^{-1}$ originally obtained from the forward Fourier-transform within R=0–3.22 Å of the first coordination shell. The fitting provided the bond distances (R) and the number of neighbors (N) for each titanium carbide material. Further information about the EXAFS fitting is provided in previous work [18].

## 3. Results and discussions

In *K*-edge X-ray absorption spectroscopy (XAS) the excitation of the Ti *1s* electron in $Ti_2CT_x$ and $Ti_3C_2T_x$ as well as in TiC, $Ti_2AlC$ and $Ti_3AlC_2$ will, as a consequence of the selection rules, only occur to molecular orbitals with *p*-character (electric dipole transitions) and *d*-character (electric quadrupole transitions). The probability of the *1s* → *3d* transitions are about one thousandth of the *1s* → *np* transitions [27,28] and the latter will therefore dominate the Ti *1s* XAS spectra. The obtained high-resolution Ti *1s* XANES spectra of TiC, $Ti_2CT_x$, $Ti_2AlC$, $Ti_3C_2T_x$, and $Ti_3AlC_2$, presented in Figure 1, show that the absorption spectra consist of two regions: the pre-edge region at 4.965-4.980 keV and the main edge region at 4.980-5.015 keV. The pre-edge features in the *K*-edge XANES spectra, peaks A and B, are assigned to the electric dipole excitation of the Ti *1s* electron into the *p-d* hybridized molecular orbitals near the Fermi energy ($E_f$), which is located at 4.9642 keV [18]. Quadrupole (*1s → 3d*) transitions can also be included, although the intensity contribution is very small because of orbital symmetry restrictions [27-29]. Also the main edge region consists of two features, C and D, where the former sometimes is referred to as the white line. The broad features C and D correspond to *1s → 4p* excitations [30]. Hence, the intensity of the pre-edge features reflects mainly the Ti *1s* electron transition into empty *p*-components of the molecular orbitals that have both *p*- and *d*-character originating from the Ti *3d* – C *2p* bonding configurations while the main edge features predominantly reflect the Ti *1s* electron transition into empty Ti *4p*-components.

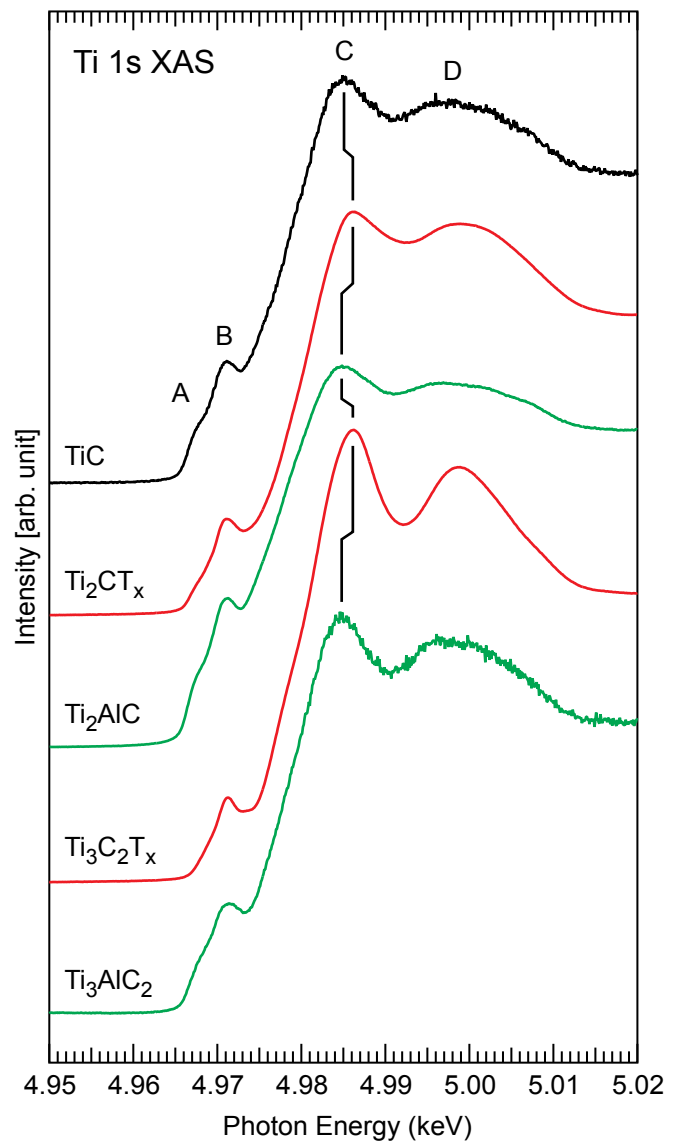

**Figure 1:** Ti *1s* XANES spectra from TiC, $Ti_2CT_x$, $Ti_2AlC$, $Ti_3C_2T_x$, and $Ti_3AlC_2$. The spectra were normalized below and above the absorption edge.

### 3.1 The Ti 1s XAS pre-edge region

The $E_f$ is located at 4.9642 keV [18] and that Figure 1 shows no features close to $E_f$ in the Ti *1s* XANES spectra of the five samples reveals that there are no electron transitions between orbitals within the probed Ti atom. Hence, all Ti *1s* electron transitions are up to unoccupied molecular orbitals where the Ti orbitals are hydridized with C orbitals, and for the MXenes also with O and F orbitals. The intensity at the pre-edge region, features A and B, originates from Ti *1s → C 2p* – Ti *3d* hybridized molecular orbitals excitations [18], where the energy separation between the two features corresponds to the crystal field splitting of the Ti *3d* states into the $t_{2g}$ and $e_g$ orbitals [30]. The pre-edge regions for the TiC, $Ti_2CT_x$, and $Ti_3C_2T_x$ samples are highlighted in Figure 2, where it is clear that the two MXene samples have their $t_{2g}$ and $e_g$ orbitals shifted further away from the $E_F$ compared to TiC. This observation suggests a larger nucleus-nucleus repulsion between C and Ti in $Ti_2CT_x$, and $Ti_3C_2T_x$, compared to TiC, which pushes the $t_{2g}$ and $e_g$ orbitals away (slightly) from the Ti nucleus [31]. Another observation is that the crystal field splitting (10Dq) for the three samples are 3.1, 2.3, and 2.2 eV. That the 10Dq is smaller for the MXene samples, compared to the TiC, suggests that the covalent Ti-C bonds are weaker in the MXene samples [31]. Further, the 10Dq is slightly reduced for the $Ti_3C_2T_x$, which indicates somewhat weaker covalent Ti-C bonds in $Ti_3C_2T_x$ compared to $Ti_2CT_x$.

The corresponding comparison in Figure 3 shows that the 10Dq for TiC, $Ti_2AlC$, and $Ti_3AlC_2$ are 3.1, 2.8, and 2.9 eV. Hence, the Ti-C bonds are strongest in TiC, although the differences in 10Dq between TiC, $Ti_2AlC$, and $Ti_3AlC_2$ are small. That 10Dq shows a small reduction between TiC and the MAX phases but a relatively larger reduction between TiC and the MXenes suggest that the redistribution of the electron clouds (free-moving electrons) caused by the electrostatic interaction with the Al-layers in $Ti_2AlC$ and $Ti_3AlC_2$ generates stronger covalent Ti-C bonds in the $Ti_3C_2$- and $Ti_2C$-layers in the MAX phases compared to the MXenes where the electron clouds are denser [23]. The influence of the Al-layers on the $Ti_3C_2$-layers in $Ti_3AlC_2$ and how it will strengthen the covalent Ti-C bonds were discussed in a recent XPS study [23]. The differences in 10Dq between the MAX phases and the MXenes, in comparison with TiC as presented in Figures 2-3, are a direct support for the discussion raised by the XPS study [23].

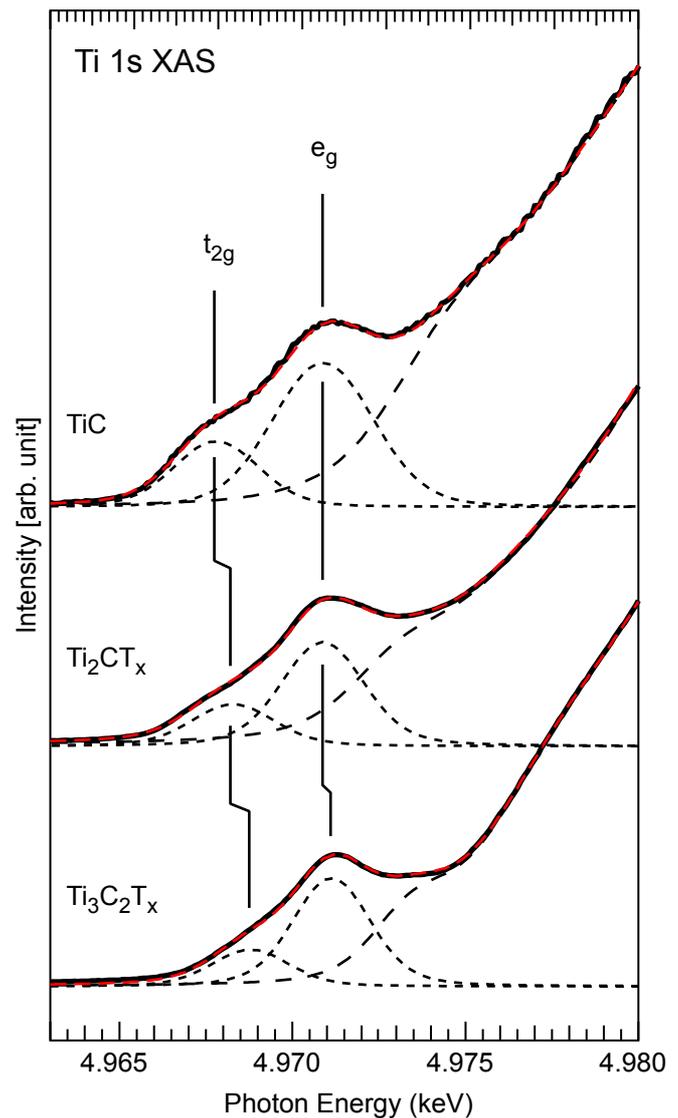

**Figure 2:** Curve fitting of the Ti *1s* XANES pre-edge region for TiC, $Ti_2CT_x$, and $Ti_3C_2T_x$, where the fitted peaks (black dotted lines) represents the Ti *1s* electron excitations into the $t_{2g}$ and $e_g$ orbitals. The fitted main edge for the Ti *1s → 4p* excitations are shown as a black dashed line. The accumulated intensity from all fitted Ti *1s* electron excitations is shown as a red dashed line.

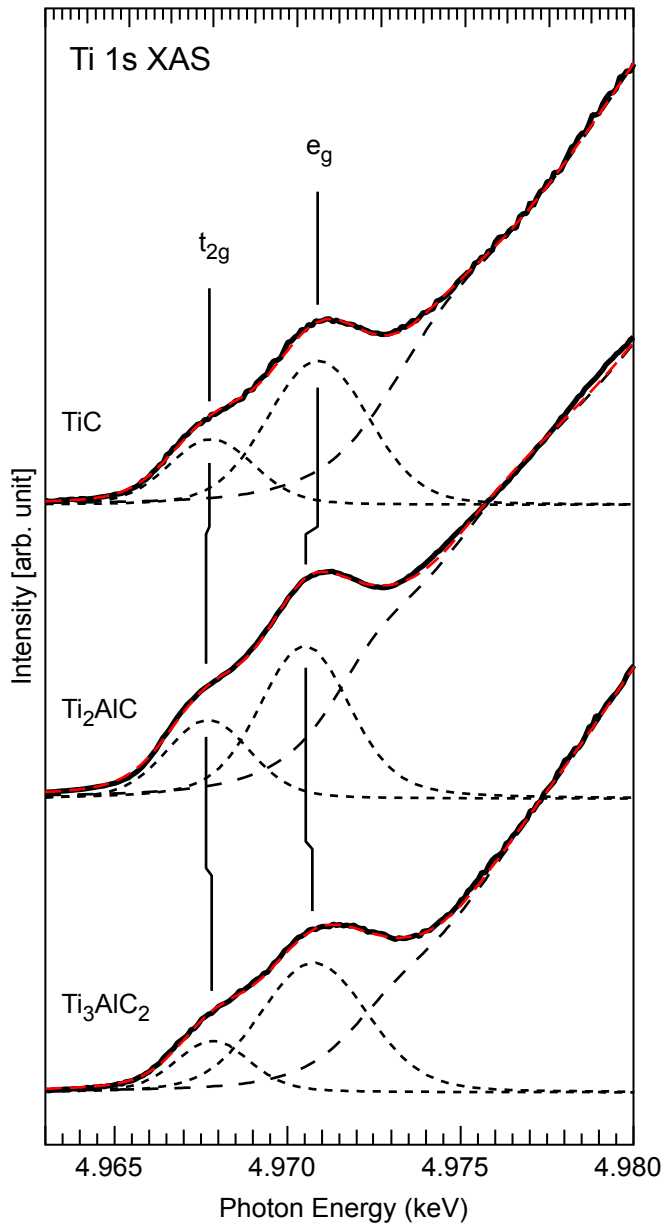

**Figure 3:** Curve fitting of the Ti *1s* XANES pre-edge region for TiC, Ti$_2$AlC, and Ti$_3$AlC$_2$, where the fitted peaks (black dotted lines) represents the Ti *1s* electron excitations into the $t_{2g}$ and $e_g$ orbitals. The fitted main edge for the Ti $\rightarrow$ *4p* excitations are shown as a black dashed line. The accumulated intensity from all fitted Ti *1s* electron excitations is shown as a red dashed line.

Figure 3 further shows that the energy positions of the $t_{2g}$ and $e_g$ orbitals for Ti$_2$AlC and Ti$_3$AlC$_2$ are about the same as for TiC, indicating that the nucleus-nucleus repulsion between C and Ti in the MAX phases differs little compared with the TiC. Hence, the C and Ti nucleus-nucleus repulsion becomes larger when the Al-layer is removed while transforming MAX phases into MXenes.

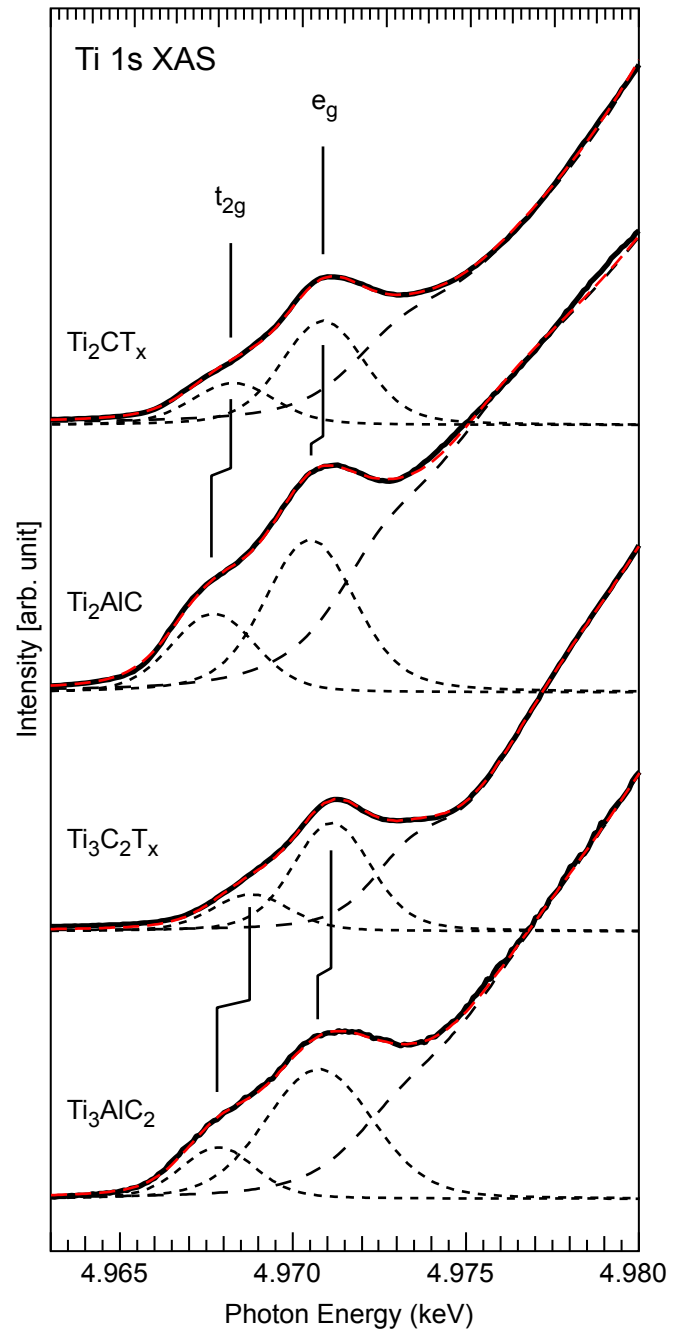

**Figure 4:** Comparison of the Ti *1s* XANES pre-edge region between the Ti$_2$CT$_x$ and Ti$_2$AlC and the Ti$_3$C$_2$T$_x$ and Ti$_3$AlC$_2$, where the fitted peaks (black dotted lines) represents the Ti *1s* electron excitations into the $t_{2g}$ and $e_g$ orbitals. The fitted main edge for the Ti $\rightarrow$ *4p* excitations are shown as a black dashed line. The accumulated intensity from all fitted Ti *1s* electron excitations is shown as a red dashed line.

It is not only the $t_{2g}$ and $e_g$ energy positions and the 10Dq that differs between the MAX phases and the MXenes. Figure 4 shows that also the intensities of the pre-edge features are dissimilar. The pre-edge intensities of the MXenes are reduced by about 40 % compared to the MAX phases. Since the intensity of the pre-edge features reflects primarily the Ti *1s* electron transition into empty p-component of the molecular orbitals that have both p- and d-character, which depends mainly on coordination, symmetry, and bond angles [32], the intensity reduction can basically be an effect from the termination species O and F that bond to Ti in $Ti_2CT_x$ and $Ti_3C_2T_x$. Bonding to O and F alters the coordination, symmetry, and bond angles for the probed Ti atoms [18,33], although most importantly is the extra electrons added to the $t_{2g}$ and $e_g$ orbitals as O and F have two and three electrons more than C; the electron configuration is [He] $2s^2 2p^4$ for O and [He] $2s^2 2p^5$ for F while it is [He] $2s^2 2p^2$ for C. The pre-edge intensity reductions, observed when the MAX phases are transformed into MXenes, can therefore be expected.

### 3.2 The Ti 1s XAS main edge region

Figure 1 shows that peak C (the white line) in the main edge region is located at 4.9850, 4.9848, and 4.9847 keV for TiC, $Ti_2AlC$, and $Ti_3AlC_2$, respectively, while both $Ti_2CT_x$ and $Ti_3C_2T_x$ have the C-peak shifted to 4.9861 keV. That TiC, $Ti_2AlC$, and $Ti_3AlC_2$ show almost the same position of peak C implies a very weak interaction between the Al-layers and the $Ti_2C$-layers in $Ti_2AlC$ and the $Ti_3C_2$-layers in $Ti_3AlC_2$, as proven in a previous XPS study [23]. Replacing the weakly interacting Al-layers with the stronger interacting termination species O and F will, on the other hand, induce significant shifts of peak C. That the C-peak has the same energy position for both $Ti_2CT_x$ and $Ti_3C_2T_x$ suggests that the Ti oxidation states are the same in $Ti_2CT_x$ and $Ti_3C_2T_x$ or that the Ti *1s* XANES shows no noticeable sensitivity towards Ti oxidation state variations in MXenes. Comparison of Ti *2p* XPS spectra of $Ti_2CT_x$ and $Ti_3C_2T_x$ supports the former [20].

The shape of the D feature is significantly different in $Ti_2CT_x$ and $Ti_3C_2T_x$ compared to TiC, $Ti_2AlC$, and $Ti_3AlC_2$. The D-peak energy position for both $Ti_2CT_x$ and $Ti_3C_2T_x$ is 4.9986 keV. However, the $Ti_3C_2T_x$ sample has one relatively sharp D-peak while the $Ti_2CT_x$ sample shows a broader feature. The TiC and the MAX phases, on the other hand, show more structures in the D-peak region. The first D-feature is located at 4.9962 keV for all three samples and in addition there are fine structures at higher energies. The Ti *1s* XANES spectrum of $TiO_2$ shows a broad D-feature with fine structures and with strong peak intensity at 5.0035 keV [30,34,35]. The previous XPS study of TiC, $Ti_3AlC_2$, and $Ti_3C_2T_x$ showed that the first two compounds are prone to oxidation while $Ti_3C_2T_x$ demonstrate more resistance against forming $TiO_2$ [23]. It is, thus, reasonable to deduce that the broad D-peak with fine structure in TiC, $Ti_2AlC$, and $Ti_3AlC_2$ originates not only from these compounds but also from oxidized material, i.e., from $TiO_2$ on the surfaces of the particles that were cold pressed into ~500 μm thick pellets. The intensity of the fine structure, especially at 5.0035 keV, indicates that the amounts of $TiO_2$ in the TiC, $Ti_3AlC_2$, and $Ti_3C_2T_x$ samples are very small. The XANES spectra of the $Ti_2CT_x$ and $Ti_3C_2T_x$ samples appear to be free from $TiO_2$-contribution.

### 3.3 The Ti 1s XAS main edge shift

A main-edge shift is often caused by changes in the charge redistribution of the probed element. When the shift is toward higher energies the charge transfer is away from the probed atoms and vice versa. In the previous XPS study the Ti *2p* XPS spectra of TiC, $Ti_3AlC_2$, and $Ti_3C_2T_x$ showed that the Ti $2p_{3/2}$ positions are 454.85, 454.75, and 455.06 eV, respectively [23]. Hence, the shift of the Ti *1s* XAS main edge (peak C) of -0.3 eV between TiC and $Ti_3AlC_2$ corresponds well with the small shift in the Ti $2p_{3/2}$ XPS, which is -0.10 eV. The corresponding shift of the Ti *1s* XAS main edge of 1.4 eV between $Ti_3AlC_2$ and $Ti_3C_2T_x$ is, on the other hand, significantly larger compared to the shift in the Ti $2p_{3/2}$ XPS, which is 0.31 eV. Hence, the shift of the Ti *1s* XAS main edge is not consistent with the charge redistribution when the MAX phase material is transformed into MXene. The shift of the Ti *1s* XAS main edge between $Ti_3AlC_2$ and $Ti_3C_2T_x$ is actually 50 % larger than expected based on the indicated charge transfer derived from the corresponding comparison of the Ti $2p_{3/2}$ XPS. Hence, the observed shift in the Ti *1s* XAS main edge (peak C), when the MAX phase material is transformed into MXene, cannot be solely derived from alteration in the oxidation state of the probed Ti atoms.

To learn more about the Ti *1s* XAS main edge shift for the $Ti_3C_2T_x$ the sample was exposed to a heating process in a protective $N_2$ atmosphere with constant gas flow. A heating process above 550 °C will desorb F from the $Ti_3C_2$ surface and the empty fcc-sites will be filled by O migrating from the Ti-Ti bridge sites, which was shown in a combined temperature-programmed XPS/HRTEM study [36]. How the heat treatment affected the Ti *1s* XANES spectrum was presented recently [18]. The detailed analysis showed that the heat treatment at 750 °C shifted the main edge by 0.5 eV. Figure 5 shows a slight shift towards higher energies in the Ti *1s* XANES when the $Ti_3C_2T_x$ sample was heated to 550 °C and a clear shift after the 750 °C heat treatment. A shift towards higher absorption energies has been interpreted as a consequence of Ti oxidation in previous electrochemical energy storage investigations, where $Ti_3C_2T_x$ has been used as a working electrode [12-17]. However, the temperature-programmed XPS study showed that F returns charge to the Ti-atoms before leaving the surface [36]. Hence, the process of F desorption reduces the oxidation state of the Ti-atoms, i.e. there is an ongoing Ti reduction process during the heat treatment. Conclusively, in an in situ heating experiment the main edge shift of the Ti *1s* XAS should be towards lower absorption energies. Figure 5 shows in contrast a main edge shift towards higher absorption energies, i.e., in the opposite direction from expected, which infers that the assumption of a direct correlation between the Ti *1s* XAS main edge shifts and changes of the Ti oxidation state in $Ti_3C_2T_x$ is not necessarily valid.

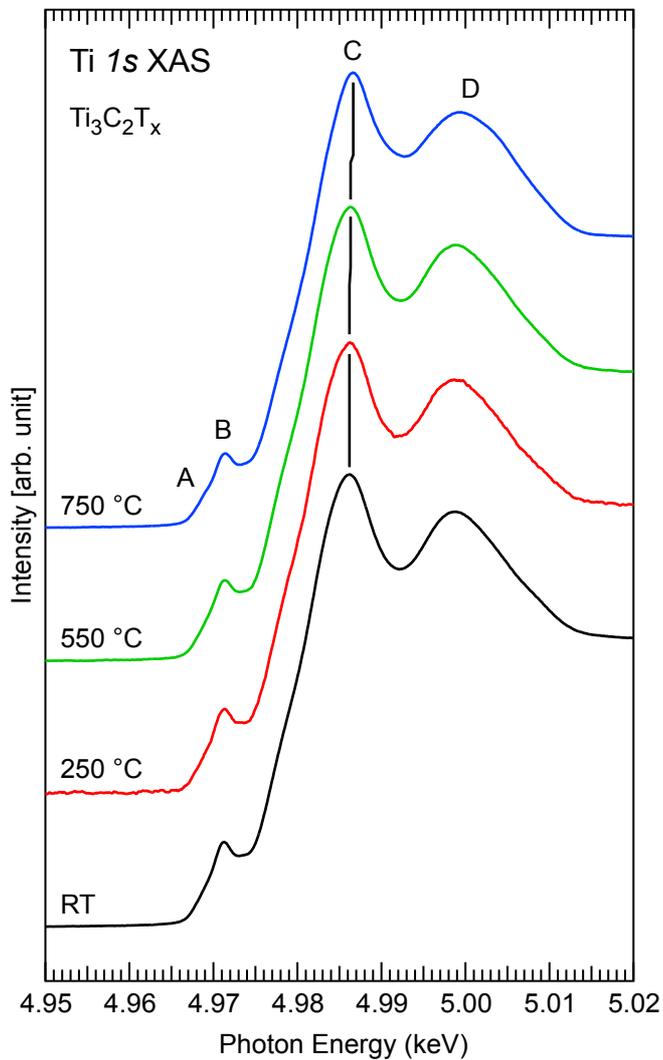

**Figure 5:** Ti *1s* XANES spectra of $Ti_3C_2T_x$ obtained at RT and after heat treatments at 250, 550, and 750 °C. No significant changes are observed after the heat treatments to 250 °C., while a small shift of the main peak (peak C) appears after the 550 °C heat treatment. The shift becomes larger after the heat treatment at 750 °C.

The Ti *1s* XAS main edge corresponds to *1s* → *4p* excitations [30] and, thus, any shifts of the main edge reflect changes in the empty Ti *4p*-components. Figure 5 shows that the Ti *1s* XANES is sensitive to changes in the bonding between the Ti and the termination species. In the in situ heating experiment we could see that the stronger bonding between the O atom and the Ti atoms around the fcc-site shifted the Ti *1s* XAS main edge toward higher energies compared to when the fcc-site is occupied by F. Thus, the observed Ti *1s* XAS main edge shift can be interpreted as an increased splitting between the occupied and the unoccupied *4p*-containing molecular orbitals, caused by the replacement of termination species at the fcc-sites. This further implies that orbitals with Ti *4p* character are involved in the bonding between the surface Ti and the termination species and that the tetragonal direction of the O orbitals leads to a stronger overlap between the O *2p* and the Ti *4p* orbitals compared with the corresponding bonding configuration between F and Ti. This explains why O remains on the $Ti_3C_2T_x$ surface while F desorbs at temperatures above 550 °C.

That a stronger orbital overlap between the O *2p* and the Ti *4p* shifts the Ti *1s* XAS main edge toward higher photon absorption energies when F is replaced by O on the fcc-sites is consistent with previous temperature-programmed XPS results [36] and a recent valence band UPS/XPS study [33].

### 3.4 The Ti 1s XAS extended region

Figure 6 shows EXAFS oscillations of structure factors $\chi(k)$ of the $Ti_2CT_x$, $T_2AlC$, $Ti_3C_2T_x$, and $Ti_3AlC_2$ displayed as a function of the wave vector k. To highlight the higher k-region, the EXAFS oscillations were $k^2$-weighted, where

$$k = \hbar^{-1} [2m(E - E_o)]^{1/2} \qquad (2)$$

corresponds to the wave vector of the excited electron in the X-ray absorption process. The horizontal arrow at the top of the figure shows the k-window for the most pronounced oscillations obtained from the raw data that has not been phase shifted. The frequency of the oscillations and the intensity of the EXAFS signal are directly related to the bond length (R) and the number of nearest neighbors (N), respectively. Higher frequency of the oscillations implies extended R while enlarged amplitude implies increased N. All structure factors show three main sharp oscillations in the 4-7 Å$^{-1}$ k-space region related to strong Ti-Ti scattering. The first main oscillation shows a low-k shift of ~0.10 Å$^{-1}$ for $Ti_2CT_x$ compared to $Ti_2AlC$, indicating shorter Ti-Ti scattering paths. A similar low-k shift of ~0.15 Å$^{-1}$ occurs also for $Ti_3C_2T_x$ compared to $Ti_3AlC_2$. These low-k shifts are attributed to relaxation of the outer Ti-layers in the MXenes compared to the MAX phases, i.e., when the attraction between the outer Ti-layers and the Al-layers vanishes the outer Ti-layers will move closer to the $M_{n+1}X_n$ core [37]. The main carbide oscillations for the MAX phases and the MXenes show correspondence to those of cubic TiC, although slightly shifted to lower k-values in comparison to TiC [18,38,39]. In addition to the main Ti-Ti peaks there are smaller peaks between them, which in some cases appear as shoulders to the main peaks. These smaller peaks and the peaks around 1-3 Å$^{-1}$ originate from Ti-C scattering. For the MXenes, the more pronounced features at low-k values at ~2.0 Å$^{-1}$ are associated with superposed oscillations from Ti-O/F scattering.

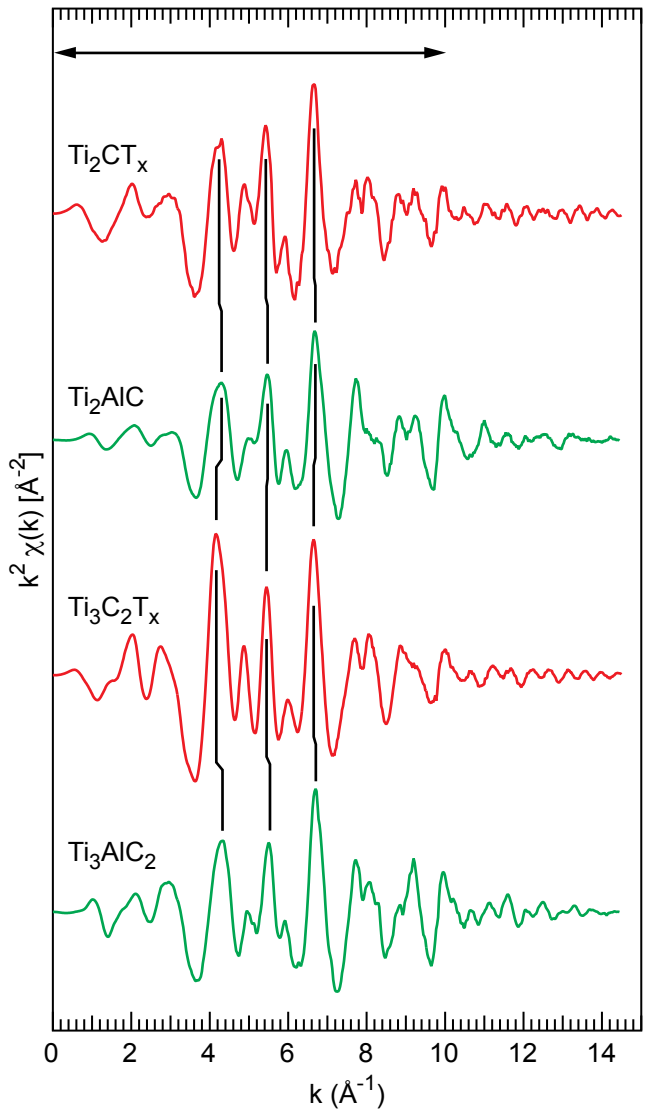

**Figure 6:** $k^2$-weighted EXAFS, $k^2\chi(k)$, as a function of the photoelectron wave number k of the $Ti_2CT_x$, $Ti_2AlC$, $Ti_3C_2T_x$, and $Ti_3AlC_2$ samples. The k-region with the most pronounced oscillations is indicated by the horizontal arrow at the top.

Figure 7 shows radial distributions functions (RDFs) of $Ti_2CT_x$, $Ti_2AlC$, $Ti_3C_2T_x$, and $T_3AlC_2$ obtained from the EXAFS oscillations in Figure 6 using Fourier transformation of the $k^2$-weighted $\chi(k)$ by the standard EXAFS procedure [40] where the radius scale has not been phase shifted. Fitting of scattering path components in the first coordination shells is indicated by dashed curves. For the MXenes the peak features between 0.5 and 3.5 Å in the first coordination shell is caused by superposed Ti-O, Ti-F, and Ti-C scattering, while for the MAX phases, the features originate only from Ti-C scattering. There are two contributions from Ti-C interactions in $Ti_3C_2T_x$ and $Ti_3AlC_2$: $Ti_I$-C for the inner Ti atoms and $Ti_{III}$-C for the outer Ti atoms. The former bonds only to C, while the latter bonds both to C and the termination species.

The three peaks in the 3-6 Å region correspond to higher coordination shells with a large contribution of long Ti-Ti scattering as they also appear in EXAFS data of cubic TiC [18,38,39]. The first peak at ~3.8 Å reflects the Ti-Ti and Ti-C scattering in the second coordination shell. The second peak at 4.5-5.2 Å reflects mainly the Ti-C-Ti and Ti-Ti-O/F scattering. The third peak at 5.5-6 Å is derived from multi-scattering paths, e.g., Ti-Ti-Ti and Ti-Ti-Ti-C. These three peaks have, in addition, a significant contribution from higher coordination shells, which contain long inclined single Ti-O scattering at 3.62, 3.73, 4.28, 4.75, 4.79, and 5.04 Å and Ti-F scattering at 3.76, 4.58, 4.78, and 4.86 Å, i.e. scattering paths from surface Ti atoms to the $T_x$ species.

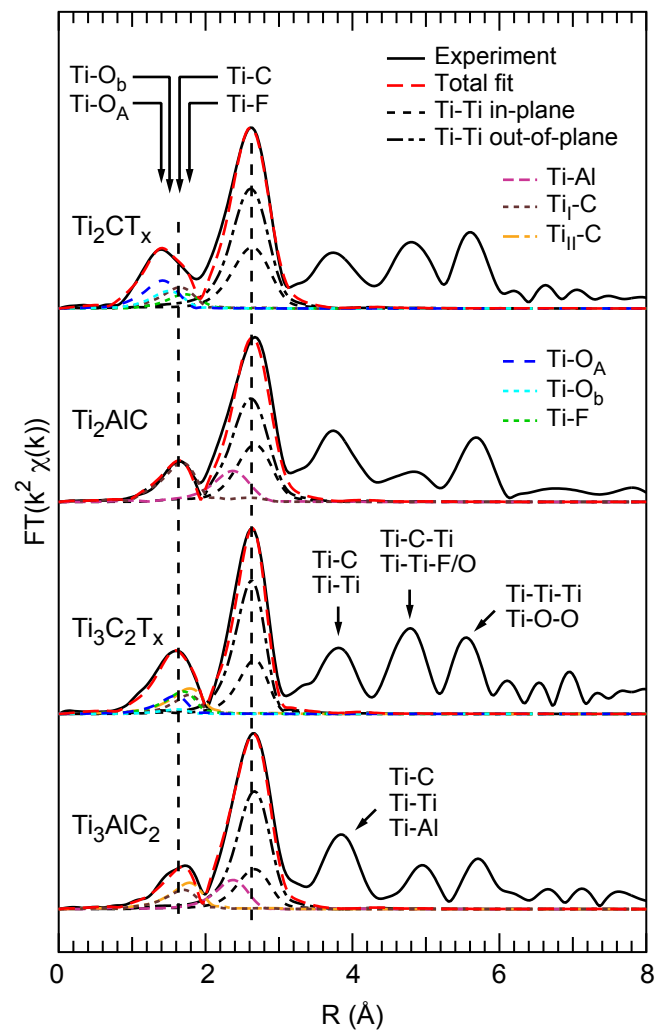

**Figure 7:** Fourier transform obtained from the $k^2$-weighted EXAFS oscillations $\chi(k)$ in Fig. 6 of $Ti_2CT_x$, $Ti_2AlC$, $Ti_3C_2T_x$, and $Ti_3AlC_2$.

Table I shows the fitting results of the EXAFS data in k- and R-space using the FEFF scattering paths of $Ti_2CT_x$,

$Ti_2AlC$, $Ti_3C_2T_x$, and $T_3AlC_2$ as model systems. For the MXenes the crystal structure modeling assumes nearly perfect $Ti_2CF_1O_2$ and $Ti_3C_2F_1O_2$ compositions in line with previous quantitative core-level XPS results [36]. The main peak of $Ti_2CT_x$ is dominated (N=4.369) by the in-plane Ti-Ti scattering at 3.001 Å and corresponds to the *a*-axis unit cell-edge in the in-plane Ti-Ti scattering. The weaker out-of-plane scattering at 3.088 Å shows a significantly lower intensity (N=2.054), which is characteristic for 2D materials.

The Ti-C bond lengths are also presented in Table 1. Although the $t_{2g}$-$e_g$ crystal field splitting is reduced for both MAX phases compared with TiC, the Ti-C bond length in $Ti_2AlC$ (2.091 Å) and $Ti_3AlC_2$ (2.210 and 2.131 Å) is slightly shorter and longer, respectively, than the Ti-C bond length in TiC (2.16 Å [41-43]). The corresponding Ti-C bond length in $Ti_2T_x$ (2.110 Å) and $Ti_3C_2T_x$ (2.209 and 2.101 Å) are showing only small adjustments compared to $Ti_2AlC$ and $Ti_3AlC_2$.

**Table I:** Fitting of the radial distribution functions in the first coordination shell provides the distance (R) and coordination number (N) for the Ti-Ti, Ti-Al and Ti-C bonds in $Ti_2CT_x$, $Ti_2AlC$, $Ti_3C_2T_x$, and $Ti_3AlC_2$. Included is also the Debye-Waller factor ($\sigma$), which is indicates the disorder and the size of the atomic displacement. The squared area of the residual is given by $\chi_r^2$ the number of independent points by $N_{ind}$, the number of fitting parameters P, and the degrees of freedom by $\nu$. The initial bond lengths in the structure models are given in parenthesis.

| System | Shell | R(Å)[a] | N[b] | $\sigma$ (Å$^2$)[c] | Statistics |
|---|---|---|---|---|---|
| $Ti_2CT_x$ | 1*Ti-O$_A$ | 1.858 | 0.409 | 0.0024 | $\chi_1^{0.95}$=3.8415 |
| | 1*Ti-O$_b$ | 1.909 | 0.256 | 0.0024 | N=28, P=27 |
| | 1*Ti-F | 2.101 | 0.235 | 0.0024 | $\nu$=N-P=1 |
| | 3*Ti-C | 2.110 (2.187) | 0.821 | 0.0021 | |
| | 6*Ti-Ti: LP[d]-*a* | 3.001 (3.040) | 4.369 | 0.0075 | |
| | 3*Ti-Ti oop[e] | 3.088 (3.145) | 2.054 | 0.0075 | |
| $Ti_2AlC$ | 3*Ti-C | 2.091 (2.109) | 0.831 | 0.0020 | $\chi_2^{0.95}$=9.586 |
| | 3*Ti-Al | 2.810 (2.838) | 0.781 | 0.0026 | N=20, P=16 |
| | 3*Ti-Ti: LP-*a* | 3.024 (3.040) | 4.669 | 0.0065 | $\nu$=N-P=4 |
| | 6*Ti-Ti oop | 3.023 (2.924) | 1.401 | 0.0065 | |
| $Ti_3C_2T_x$ | 1*Ti-O$_A$ | 1.972 | 0.511 | 0.0020 | $\chi_3^{0.95}$=7.815 |
| | 1*Ti-O$_b$ | 1.983 | 0.102 | 0.0020 | N=30, P=28 |
| | 1*Ti-F | 2.110 | 0.442 | 0.0024 | $\nu$ =N-P=3 |
| | 6*Ti$_I$-C | 2.209 (2.227) | 1.463 | 0.0021 | |
| | 3*Ti$_{II}$-C | 2.101 (2.120) | 0.975 | 0.0021 | |
| | 6*Ti-Ti: LP[d]-*a* | 3.011 (3.080) | 4.799 | 0.0065 | |
| | 3*Ti-Ti oop[e] | 3.090 (3.065) | 2.159 | 0.0065 | |
| $Ti_3AlC_2$ | 6*Ti$_I$-C | 2.210 (2.221) | 1.053 | 0.0020 | $\chi_4^{0.95}$=9.487 |
| | 3*Ti$_{II}$-C | 2.131 (2.113) | 0.701 | 0.0020 | N=24, P=20 |
| | 3*Ti$_{II}$-Al | 2.747 (2.776) | 0.887 | 0.0026 | $\nu$ =N-P=4 |
| | 6*Ti-Ti: LP[d]-*a* | 3.040 (3.060) | 4.976 | 0.0065 | |
| | 3*Ti-Ti oop[e] | 3.051 (3.066) | 1.493 | 0.0065 | |

[a] The errors of the atomic distances are estimated to be ±0.01 Å

[b] The errors in the coordination numbers are estimated to be ±0.01 Å.

[c] The errors in the Debye-Waller factors are estimated to be ±0.001 Å$^2$.

[d] Lattice parameter.

[e] Out-of-plane.

Nevertheless, for $Ti_3C_2T_x$, the Ti-C bond length of the outer Ti atoms is 0.108 Å shorter than for the inner Ti atoms, which is similar as for $Ti_3AlC_2$ where the corresponding Ti-C bond length is 0.079 Å shorter. Hence, while the Ti-C bond length of the inner Ti atoms is the same for both $Ti_3C_2T_x$ and $Ti_3AlC_2$, the Ti-C bond length of the outer Ti atoms is reduced by 0.03 Å when the Al-layer is replaced by the termination species. This finding differs from a previous EXAFS study [18], where the initial structure model used for the fitting procedure of the $Ti_3C_2T_x$ was based on a theoretical work [44]. In the present study the initial structure model of the $Ti_3C_2T_x$ was a $Ti_3C_2$-layer in $Ti_3AlC_2$ with termination species bonding condition based on a valence band XPS study [33]. Comparing the fitting result of $Ti_3C_2T_x$ in the present EXAFS study with the previous EXAFS study, i.e., when the initial structure is based on a $Ti_3C_2$-layer in $Ti_3AlC_2$ versus when the initial structure is based on a theoretical optimized $Ti_3C_2T_x$, shows that the former initial structure provides a better fit. Hence, the use of theoretical models must be performed with caution.

### 3.5 Ti-C bond strength versus Ti-C bond length

From the XANES data, we observe that the $t_{2g}$-$e_g$ crystal field splitting in $Ti_2AlC$ and $Ti_3AlC_2$ are only 10 % and 6 % lower compared to TiC, while the corresponding splitting for $Ti_2CT_x$ and $Ti_3C_2T_x$ are reduced by 26 % and 29 %. Hence, the strength of the covalent Ti-C bonds in the MAX phases is only slightly weaker compared to TiC whereas the corresponding covalent Ti-C bond strength in the MXenes is significantly weaker. The reduced $t_{2g}$-$e_g$ crystal field splitting in the MXenes, compared to the MAX phases, implies that the replacement of the Al-layers by the termination species has weakened the covalent Ti-C bonding significantly. However, the removal of the Al-layers from the MAX phases redistributes the electron cloud (the free moving electrons) and makes it denser in $Ti_2CT_x$ and $Ti_3C_2T_x$ compared with the $Ti_2C$- and $Ti_3C_2$-layers in the MAX phases [23]. The internal bonding in $Ti_2AlC$ and $Ti_3AlC_2$ as well as in $Ti_2CT_x$ and $Ti_3C_2T_x$ is a combination of covalent type of bonding and metallic type of bonding. Hence, the redistribution of the electron cloud when $Ti_2AlC$ and $Ti_3AlC_2$ are transformed into $Ti_2CT_x$ and $Ti_3C_2T_x$, respectively, enables a weakening of the covalent Ti-C bonds, which facilitates the covalent bonding to the termination species.

It is important to remember that the Ti-C bond strength variations reflected by the $t_{2g}$-$e_g$ crystal field splitting is influenced by the overlap of the C $2p$ and Ti $3d$ orbitals while the average bond length in $Ti_2CT_x$ and $Ti_3C_2T_x$ is a result of the contribution from both the covalent bonding from the orbital overlap and the electron cloud distribution. Hence, a denser electron cloud can induce an adjustment in the atomic positions leading to a longer or shorter distance between the atoms. In the case when $Ti_2AlC$ is transformed into $Ti_2CT_x$ the Ti-C bond length remains unchanged. The same is observed for the inner Ti-C bond length when $Ti_3AlC_2$ is transformed into $Ti_3C_2T_x$ and the outer Ti-C bond length is reduced by only 0.03 Å. Hence, the denser electron cloud will strengthen the metallic type of bonding making it necessary to reduce the strength of the covalent type of bonding between the Ti and C to retain the crystal structure.

In the process where the Al-layers are removed in the MAX phase to MXene transition there is a sensitive time window where the balance between the covalent type of bonding and the metallic type of bonding is disrupted. The covalent Ti-C bond may very well be weakened at that moment before the electron cloud has become stable and dense enough. Consequently, there is a significant risk of Ti-C bond breakage and inevitably a devastating formation of graphite-like components in this time window. Hence, it is crucial that the etching process that transforms MAX phases into MXenes is optimized and carefully performed to minimize the effect of the electron cloud/covalent Ti-C bond re-balancing activity or otherwise the high amount of graphite-like components can lead to failing functionality and questioned results from applied studies using MXenes.

### 3.6 The Ti 1s XAS main edge shift versus electrochemical cycling investigations

That the Ti $1s$ XAS main edge shift is larger than expected for the MXene samples and that the replacement of F by O in the fcc-sites shifts the main edge towards higher photon absorption energies raises the question: is it appropriate to assume that the origin of an observed shift is because of Ti oxidation? The question is relevant as recently presented electrochemical cycling investigations have assumed that the observed shifts of the Ti $1s$ XAS main edge are simply because of changes in the Ti oxidation state of $Ti_3C_2T_x$ when the applied potential over the electrochemical cell alternates from low to high potentials and reverse [12-17]. Although the present Ti $1s$ XAS study cannot exclude a possible oxidation of the Ti atoms in the electrochemical cycling investigations the conclusions from the present work suggest that the observed main edge shift when the applied potential over the electrochemical cell changes from low to high potential is predominantly because of modifications of the termination species at the fcc-sites. Hence, the observed shifts in the electrochemical cycling investigations are because of the response from the adsorption and desorption processes on the termination species coordinated to the fcc-sites.

One might argue that heat treatment and electrochemical cycling are two different experiments involving two different processes. However, the Ti $1s$ XANES of the two experiments show the same size of main edge shift in the same direction, which indicates that the effect on the Ti atoms is the same in these two experiments. Our work shows that it is a desorption process involving the fcc-sites that causes the observed Ti $1s$ XANES main edge shift toward higher photon absorption energies, even though the Ti $1s$ XANES main edge should shift toward lower photon absorption energies because of a charge transfer to the Ti atoms from the desorbing F atoms. Hence, the Ti $1s$ XANES main edge position is more sensitive to adsorption and desorption processes involving the termination species coordinated to the fcc-sites on the MXenes than small changes in the Ti oxidation state. In previous interpretations of the observed main edge shifts toward higher energies in combined XAS/electrochemical cycling studies [12–17] the interpretation is that Ti loses charge in an oxidation process. However, our work shows that the observed main edge shifts in combined XAS/electrochemical cycling are instead because of adsorption and desorption processes on the termination species that will alter the charge of the Ti around the fcc-sites.

Periodical intercalation of ions between 2D MXene layers in electrochemical cycling processes has shown encouraging results regarding applications in energy storage devices where ions are the charge carriers, such as capacitors, batteries, and hydride electrochemical devices [10]. However, the development of such devices requires a correct description of the intercalation process and the roles of the $T_x$ and the carbide core layers. The Ti $1s$ XAS study presented in this work provides essential information about the structure, bond strength, and active sites for the intercalation processes that brings us closer to a complete understanding of MXenes and their unique characteristics important for future energy storage devices.

## 4. Conclusions

Through high-resolution Ti $1s$ XAS it was possible to perform near edge absorption studies as well as extended region oscillations studies of high quality $Ti_3C_2T_x$, $Ti_3AlC_2$, $Ti_2CT_x$, and $T_2AlC$ samples. From the XANES we conclude that the crystal field splitting (10Dq) of the $t_{2g}$ and $e_g$ orbitals, observed in the pre-edge region of the XANES, reveals a reduced Ti $3d$ – C $2p$ orbital overlap in the MXenes compared to the MAX phases, which suggests a weakening of the covalent Ti-C bond strength when the MAX phases are transformed into MXenes. However, the small differences in bond lengths between the MXenes and the MAX phases, as confirmed through EXAFS analysis, indicate that the weakening of the covalent Ti-C bond strength are compensated by an increase of the metal bonding in the $Ti_{n+1}C_n$-layers. The extended XAS region further shows that the Ti-C bonds are shorter in $Ti_2CT_x$, and $T_2AlC$ compared to $Ti_3C_2T_x$ and $Ti_3AlC_2$.

Moreover, according to the $t_{2g}$-$e_g$ crystal field splitting the Ti-C bond strength in the carbide core is slightly stronger for $Ti_3AlC_2$ than for $Ti_2AlC$, most likely because with five monolayers (three Ti-monolayers and two C-monolayers) in $Ti_3AlC_2$ there are better conditions for adjustments of the atomic positions for more efficient orbital overlaps between the Ti $3d$ orbitals and the C $2p$ orbitals. The different electron cloud distribution in $Ti_3AlC_2$ and $Ti_2AlC$ is also playing an important role [23], where the Al-monolayer in $Ti_2AlC$ attracts more charge compared to $Ti_3AlC_2$.

From the EXAFS we reveal that the Ti-C distances remain intact in $Ti_2CT_x$ and in the inner Ti-C distances in $Ti_3C_2T_x$ when the Al-layers are removed from $Ti_2AlC$ and $Ti_3AlC_2$, respectively, while the outer Ti-layers in $Ti_3C_2T_x$ relaxes and move slightly closer to the C-layers. In addition, the results obtained through EXAFS support the bond lengths between the surface Ti-atoms and the termination species O and F in $Ti_3C_2T_x$ as concluded in a previous valence band XPS study [33].

We also conclude that the Ti $4p$ orbital is involved in the bonding between the surface Ti and the termination species in the MXenes. The Ti $4p$ involvement makes the Ti $1s$ XAS main edge sensitive toward the termination species and potential adsorption and desorption processes on the termination species in the fcc-sites. Ti $1s$ XAS is, thus, a suitable tool in intercalation process studies of energy storage devices. Consequently, the Ti $1s$ XANES main edge shifts observed in electrochemical cycling investigations of MXenes [12–17] are because of adsorption and desorption processes on the termination species coordinated to the fcc-sites on the MXenes.

## Acknowledgements

We thank Dr. Joseph Halim for preparing the samples and the staff at the MAX IV Laboratory for experimental support. Research conducted at MAX IV, a Swedish national user facility, is supported by the Swedish Research council under contract 2018-07152, the Swedish Governmental Agency for Innovation Systems under contract 2018-04969, and Formas under contract 2019-02496. We also thank the Swedish Government Strategic Research Area in Materials Science on Functional Materials at Linköping University (Faculty Grant SFO-Mat-LiU No. 2009 00971). M.M. acknowledges financial support from the Swedish Energy Research (Grant No. 43606-1) and the Carl Tryggers Foundation (CTS20:272, CTS16:303, CTS14:310). The FEFF calculations of scattering paths were performed using supercomputer resources provided by the National Academic Infrastructure for Supercomputing in Sweden (NAISS) at Linköping University funded by the Swedish Research Council through grant agreement no. 2022-06725.

## References

[1]  Ghidiu M, Lukatskaya M R, Zhao M-Q, Gogotsi Y and Barsoum M W 2014 *Nature* **516** 78-81
[2]  Naguib M, Halim J, Lu J, Cook K M, Hultman L, Gogotsi Y and Barsoum M W 2013 *J. Am. Chem. Soc.* **135** 15966–15969
[3]  Li N, Chen X, Ong W-J, MacFarlane D R, Zaho X, Cheetham A K and Sun C 2017 *ACS Nano* **11** 10825-10833
[4]  Halim J, Lukatskaya M R, Cook K M, Lu J, Smith C R, Näslund L-Å, May S J, Hultman L, Gogotsi Y, Eklund P and Barsoum M W 2014 *Chem. Mater.* **26** 2374-2381
[5]  Persson I, Halim J, Lind H, Hansen T W, Wagner J B, Näslund L-Å, Darakchieva V, Palisaitis J, Rosen J and Persson P O Å 2019 *Adv. Mater.* **31** 1805472
[6]  Li X, Yin X, Song C, Han M, Xu H, Duan W, Cheng L and Zhang L 2018 *Adv. Funct. Mater.* **28** 1803938
[7]  Lukatskaya M R, Kota S, Lin Z, Zhao M-Q, Shpigel N, Levi M D, Halim J, Taberna P-L, Barsoum M W, Simon P and Gogotsi Y 2017 *Nat. Energy* **2** 17105
[8]  Naguib M, Come J, Dyatkin B, Presser V, Taberna P L, Simon P, Barsoum M W and Gogotsi Y, 2012 *Electrochem. Commun.* **16** 61-64
[9]  Gentile A, Ferrara C, Tosoni S, Balordi M, Marchionna S, Cernuschi F, Kim M H, Lee H W and Ruffo R 2020 *Small Methods* **4** 2000314
[10]  Lukatskaya M R, Mashtalir O, Ren C E, Dall'Agnese Y, Rozier P, Taberna P L, Naguib M, Simon P, Barsoum M W and Gogotsi Y 2013 Science **341** 1502-1505
[11]  Aslam M K, Niu Y and Xu M 2021 *Adv. Energy Mater.* **11** 2000681
[12]  Lukatskaya M R, Bak S-M, Yu X, Yang X-Q, Barsoum M W and Gogotsi Y 2015 *Adv. Energy Mater.* **5** 1500589
[13]  Xie Y, Naguib M, Mochalin V N, Barsoum M W, Gogotsi Y, Yu X, Nam K-W, Yang X-Q, Kolesnikov A I and Kent P R C 2014 *J. Am. Chem. Soc.* **136** 6385-6394
[14]  Wang X, Wang S, Qin J, Xie X, Yang R and Cao M 2019 *Inorg. Chem.* **58** 16524-16536
[15]  Wang X, Wang J, Qin J, Xie X, Yang R and Cao M 2020 *ACS Appl. Mater. Interfaces* **12** 39181-39194
[16]  Ferrara C, Gentile A, Marchionna S, Quinzeni I, Fracchia M, Ghigna P, Pollastri S, Ritter C, Vanacore G M and Ruffo R 2021 *Nano Lett.* **21** 8290-8297
[17]  Wang X, Bak S-M, Han M, Shuck C E, McHugh C, Li K, Li J, Tang J and Gogotsi Y 2022 *ACS Energy Lett.* **7** 30-35
[18]  Magnuson M and Näslund L-Å 2020 *Phys. Rev. Research* **2** 033516
[19]  Naguib M, Kurtoglu M, Presser V, Lu J, Niu J, Heon M, Hultman L, Gogotsi Y and Barsoum M W 2011 *Adv. Mater.* **23** 4248-4253
[20]  Natu V, Benchakar M, Canaff C, Habrioux A, Célérier S and Barsoum M W 2021 *Matter* **4** 1224–1251
[21]  Bordwehr V S R 1989 *Am. Phys. Fr.* **14** 377-466
[22]  Magnuson M and Mattesini M 2017 *Thin Solid Films* **621** 108–130
[23]  Näslund L-Å, Persson P O Å and Rosen J 2020 *J. Phys. Chem. C* **124** 27732-27742
[24]  Klementev K V 2001 *J. Phys. D. Appl. Phys.* **34** 209–217
[25]  Rehr J J, Kas J J, Vila F D, Prange M P and Jorissen K 2010 *Phys. Chem. Chem. Phys.* **12** 5503–5513

[26]  Rehr J J, Kas J J, Prange M P, Sorini A P, Takimoto Y and Vila F D 2009 *C. R. Phys.* **10** 548–559
[27]  Bair R A and Goddard III W A 1980 *Phys. Rev. B* **22** 2767-2776
[28]  Kawai J 2000 *Absorption techniques in X-ray spectroscopy. In Encyclopedia of Analytical Chemistry* ed R A Meyers (Wiley, Chichester) p 13288
[29]  Yamamoto T 2008 *X-Ray Spectrom.* **37** 572–584
[30]  Triana C A, Araujo C M, Ahuja R, Niklasson G A and Edvinsson T 2016 *Phys. Rev. B* **94** 165129
[31]  Atkins P W 1983 *Molecular Quantum Mechanics* 2nd edn (Oxford: Oxford University Press, Inc.)
[32]  Jiang N, Su D and Spence J C H 2007 *Phys. Rev. B* **76** 214117
[33]  Näslund L-Å, Mikkelä M-H, Kokkonen E and Magnuson M 2021 *2D Mater.* **8** 045026
[34]  Farges F, Brown Jr. G E and Rehr J J 1997 *Phys. Rev. B* **56** 1809-1819
[35]  Rossi T C, Grolimund D, Nachtegaal M, Cannelli O, Mancini G F, Bacellar C, Kinschel D, Rouxel J R, Ohannessian N, Pergolesi D, Lippert T and Chergui M 2019 *Phys. Rev. B* **100** 245207
[36]  Persson I, Näslund L-Å, Halim J, Barsoum M W, Darakchieva V, Palisaitis J, Rosen J and Persson P O Å 2017 *2D Mater.* **5** 015002
[37]  Zangwill A 1988 *Physics at Surfaces* (New York: Cambridge University Press)
[38]  Balzarotti A, De Crescenzi M and Incoccia L 1982 *Phys. Rev. B* **25** 6349-6366
[39]  Adelhelm C, Balden M and Sikora M 2007 *Materials Sci. Eng. C* **27** 1423-1427
[40]  Rehr J J and Albers R C 2000 *Rev. Mod. Phys.* **72** 621–654
[41]  Storms E K 1967 *The Refractory Carbides* (Academic Press, New York)
[42]  Storms E K and McNeal R J 1962 *J. Phys. Chem.* **66** 1401-1408
[43]  Sanjinés R, Wiemer C, Hones P and Lévy F 1998 *J. Appl. Phys.* **83** 1396-1402
[44]  Khazaei M, Arai M, Sasaki T, Chung C-Y, Venkataramanan N S, Estili M, Sakka Y and Kawazoe, Y 2013 *Adv. Funct. Mater.* **23** 2185–2192